\documentclass[letterpaper, 10 pt, conference]{ieeeconf} 
\IEEEoverridecommandlockouts
\usepackage{cite}
\usepackage{amsmath,amssymb,amsfonts}
\usepackage{algorithmic}
\usepackage{graphicx}
\usepackage{textcomp}
\usepackage{xcolor}
\usepackage{url}
\usepackage[final]{listings}
\usepackage{listing-td}
\usepackage{amsmath}
\usepackage{empheq}
\usepackage{multirow}

\usepackage{xcolor} 
\definecolor{LightGray}{rgb}{0.97,0.97,0.97}
\definecolor{Sepia}{rgb}{0.8,0.3,0}
\definecolor{Green}{rgb}{0.0,0.4,0}
\definecolor{Purple}{rgb}{0.5,0,0.9}
\definecolor{MidnightBlue}{rgb}{0.0,0,0.97}

\usepackage{hyperref}
\usepackage{listings}
\usepackage{amsmath}
\usepackage{lipsum}
\usepackage{graphics}
\usepackage{array}
\usepackage{mdframed}

\definecolor{eqgray}{RGB}{230,230,230}

\usepackage[normalem]{ulem}
\useunder{\uline}{\ul}{}
\lstdefinelanguage{SPARQL}{
  basicstyle=\small\ttfamily,
  backgroundcolor=\color{LightGray},
  columns=fullflexible,
  breaklines=false,
  sensitive=true,
  frame=bt,
  aboveskip=1em,
  belowskip=1em,
  xleftmargin=.5em,
  xrightmargin=.5em,
  framexleftmargin=.5em,
  framextopmargin=.5em,
  framexbottommargin=.5em,
  framexrightmargin=.5em,
  tabsize = 2,
  showstringspaces=false,
  morecomment=[l][\color{gray}]{\#},       
  morecomment=[n][\color{blue}]{<http}{>}, 
  morestring=[b][\color{OliveGreen}]{\"},  
  keywordsprefix=?,
  classoffset=0,
  keywordstyle=\color{Sepia},
  morekeywords={},
  classoffset=1,
  keywordstyle=\color{Purple},
  morekeywords={rdf,rdfs,owl,xsd,purl,phy,req,sys,td,pdt,hvac,brick},
  classoffset=2,
  keywordstyle=\color{Green},
  morekeywords={eka},
  classoffset=1,
  keywordstyle=\color{blue},
  morekeywords={System},
  classoffset=2,
  keywordstyle=\color{Purple},
  morekeywords={ControlProgram},  
  classoffset=3,
  keywordstyle=\color{MidnightBlue},
  morekeywords={
    SELECT,CONSTRUCT,DESCRIBE,ASK,WHERE,FROM,NAMED,PREFIX,BASE,OPTIONAL,
    FILTER,GRAPH,LIMIT,OFFSET,SERVICE,UNION,EXISTS,NOT,BINDINGS,MINUS,a
  }
}

\def\BibTeX{{\rm B\kern-.05em{\sc i\kern-.025em b}\kern-.08em
    T\kern-.1667em\lower.7ex\hbox{E}\kern-.125emX}}
\begin{document}

\title{A Match Made in Semantics: Physics-infused Digital Twins for Smart Building Automation}

\author{\authorblockN{Ganesh Ramanathan}
\authorblockA{\textit{Siemens AG} \\
Zug, Switzerland \\
ganesh.ramanathan@siemens.com}
\and
\authorblockN{Simon Mayer}
\authorblockA{\textit{University of St. Gallen} \\
St.Gallen, Switzerland \\
simon.mayer@unisg.ch}
}

\maketitle


\begin{abstract}
Buildings contain electro-mechanical systems that ensure the occupants' comfort, health, and safety.
The functioning of these systems is automated through control programs, which are often available as reusable artifacts in a software library.
However, matching these reusable control programs to the installed technical systems requires manual effort and adds engineering cost.
In this article, we show that such matching can be accomplished fully automatically through logical rules and based on the creation of semantic relationships between descriptions of \emph{physical processes} and descriptions of technical systems and control programs.
For this purpose, we propose a high-level bridging ontology that enables the desired rule-based matching and equips digital twins of the technical systems with the required knowledge about the underlying physical processes in a self-contained manner.
We evaluated our approach in a real-life building automation project with a total of 34 deployed air handling units. Our data show that rules based on our bridging ontology enabled the system to infer the suitable choice of control programs automatically in more than 90\% of the cases while avoiding almost an hour of manual work for each such match.
\end{abstract}


\section{Introduction}

Buildings contain systems (e.g., for heating, ventilation, electrical energy management, plumbing, etc.) that comprise sub-systems and components (e.g., boilers, pumps, fans, etc.), each of which manages physical mechanisms that are involved in the transformation of substance or energy (i.e., physical \textit{stuff}~\cite{pp_borst1995physsys}).
Domain experts have developed libraries of reusable control programs for automating many types of systems commonly found in buildings.
In many cases, this avoids the automation engineer having to develop (and test) control strategies from scratch.
For example, a control program for a room's heating system can be developed once and reused across multiple installations because the system's technical components and topology are similar and the underlying physical processes are the same~\cite{bamod_ramanathan2023reasoning}.

However, the automation engineer today still has to manually read and understand the descriptions of the control programs available in the library and also the design documentation of the technical system in order to select a suitable program to reuse---i.e., they need to match specific system designs with control programs.
An \emph{automated} way to select the correct control program and adapt its configuration to a technical system would reduce the cost and effort required for such re-engineering~\cite{bamod_butzin_engineering,engineering_dibowski_integrated}.

Researchers in both process engineering and automation have developed several ontologies to create machine-understandable knowledge about technical systems, control programs, and physical processes.
Several authors have hinted that an integration of these might enable automated matching of a technical system to a suitable control program~\cite{engineering_runde2010engineering,as_schneider2017ontology,bamod_delgoshaei2022semantic,bamod_ramanathan2023reasoning}.
%

However, the concepts that are required for interlinking the individual abstractions have not been fully explored and the engineering ontologies currently available in the building automation (BA) domain do not address this need.
On the other hand, establishing a new set of ontologies for this purpose is difficult to achieve and cumbersome to maintain considering the multi-vendor and fragmented landscape in automation systems, including in BA~\cite{bamod_butzin_engineering,engineering_challenges_vogel2014challenges}. This setup with multiple heterogeneous vendors also poses the challenge of achieving a centralized knowledge base since it requires coordination and collaboration for data ownership and governance.

In this paper, we argue and demonstrate that both these challenges can be conceptually overcome through a \emph{high-level ontology} that coordinates (but does not replace) vendor ontologies.
Our approach is applicable in the context of BA as well as in other domains with similar challenges.
By providing bridging concepts, the high-level ontology enables automated matching and, by its ability to integrate knowledge, also enables the creation of self-contained descriptions of technical systems and control programs, which can be embedded into digital twins, thereby achieving a \emph{decentralized knowledge repository}.
Our core contributions are:
\begin{enumerate}
    \item We report on the results of our study of the existing ways control programs, technical systems, and physical processes are described and further explain that the knowledge of the underlying physical processes is essential for reasoning about the suitability of a control program to a technical system towards automatic matching.
    
    \item We propose a high-level ontology that can bridge concepts in the existing ontologies for modeling technical systems and physical processes.
    This helps designers of both technical systems and control programs to include the knowledge of the physical process in their (machine-understandable) design descriptions, which enables creation of generic semantic rules for matching.
    
    \item We introduce the World Wide Web Consortium's Web of Things \textit{Thing Description} standard as a way to embed machine-understandable descriptions of the physical processes into a standardized and interoperable representation of the \textit{digital twin} of the technical system. Such \textit{physics-infused digital twins} enable the rule-based matching to operate on knowledge that is accessible in a distributed and self-contained manner.  

    \item We show the results of evaluating our approach in a real-life BA system modeled using currently available engineering ontologies. Our results confirm that our approach permits the automatic matching of technical systems and control programs. Furthermore, we discuss which aspects of our bridging concepts were particularly relevant for automated matching.
\end{enumerate}



\section{Related Work}
\label{sec:relwork}

This section introduces existing work that is relevant to integrating knowledge towards enabling automated selection or configuration of control programs, and highlights the gaps which we addressed in our approach.

\subsection{Technical Systems}

Design descriptions of technical systems (TS) are traditionally created in human-readable form only (e.g., CAD drawings, texts, etc.). In recent years, there has been an increased use of Semantic Web technologies, which allows experts to create reusable ontologies that capture \textit{machine-understandable} knowledge about their specific domains.
Over the past decade, the Semantic Web has propelled the creation of ontologies in the building systems domain too~\cite{swt_industrial_biffl2016semantic,swt_pauwels2017semantic} .
For example, ontologies such as BRICK~\cite{sd_balaji2016brick}, Haystack~\footnote{\url{https://project-haystack.org/doc/lib-phIoT}}, RealEstateCore~\footnote{\url{https://www.realestatecore.io/}}, SAREF4BLDG~\footnote{\url{https://saref.etsi.org/saref4bldg/v1.1.2/}} (for building systems), SSN/SOSA\footnote{\url{https://www.w3.org/TR/vocab-ssn/}} (for sensors and actuators), and Tubes~\cite{sd_pauen2021tubes}(system design) can be used to describe the sub-systems, components, and their inter-dependencies.
The principal language used for this purpose is the Resource Description Framework (RDF) and its extension Web Ontology Language (OWL)~\footnote{\url{https://www.w3.org/OWL/}} which is based on a decidable subset of first-order logic and can hence be used for automated reasoning.
A similar approach for model-based engineering, though not based on the Semantic Web, is offered by the AutomationML standard\footnote{\url{https://www.automationml.org/}} which is widely used in the factory and process automation domains; ~\cite{aml_kovalenko2018automationml} proposes an OWL ontology for this standard.

Such machine-understandable system descriptions can be used during both design~\cite{bamod_butzin_engineering,engineering_dibowski_integrated} and operation~\cite{bamod_ontology_ramanathan,bamod_delgoshaei2022semantic} of an automation system.
Further,~\cite{bamod_ramanathan2023reasoning,as_schneider2017ontology} show that structural descriptions can be extended with knowledge about the physical interfaces of components and their relationship with the underlying physical processes.

\subsection{Control Programs}

Several standards like IEC 61131 and its implementation in PLCOpen\footnote{\url{https://plcopen.org/}}, CIF\footnote{\url{https://eclipse.dev/escet/v2.0/cif/}}, Control Description Language (CDL)\footnote{\url{https://obc.lbl.gov/specification/cdl.html}} etc. exist to support the reusability of control programs (CPs) for common types of systems.
On this basis, ontologies have been proposed to permit machines to understand the purpose of a CP and the use of its external interfaces---prominent amongst these are CTRLOnt ~\cite{as_schneider2017ontology} and a semantic extension to CDL described in~\cite{as_wetter2019control}.

However, current approaches to semantically describing CPs remain limited to annotating their IO interfaces with information about the physical quantity, measurement units, and type of \emph{stuff} they deal with.
We propose to in addition formalize the relationship of CPs with TSs and their underlying physical processes. To this end, we suggest a high-level ontology which allows the designer of a CP to relate its external interfaces to both: interfaces of the TS and variables of the underlying physical process. 

\subsection{Physical Processes and Mechanisms}
Physical mechanisms like energy conversion, heat exchange, or fluid flow are modeled using mathematical equations which are based on laws in physics that describe the static and dynamic behavior.
Such mechanisms can be coupled to each other through the flow of substance or energy (often referred to as physical \textit{stuff}) to form a physical process.
The process engineering community has developed several ontologies to semantically describe physical processes in terms of connected physical mechanisms and the variables which represents it state.
We will use the term \textit{physical process description} (PPD) to denote this knowledge.
This knowledge is essential because control programs are often designed taking into consideration the type and composition of the physical mechanisms involved in a process~\cite{pm_lee2008CPSDesign,pp_morbach2009ontocape,pp_kief_yoshioka2004physical}.
Prominent amongst the ontologies to model PPDs are OntoCAPE (in OWL)~\cite{pp_morbach2009ontocape} and PhySys~\cite{pp_borst1995physsys}.
Ontologies like Qudt\footnote{\url{https://www.qudt.org/}} and Scientific Variables Ontology\footnote{\url{https://scientificvariablesontology.org/}} provide an extensive vocabulary of physical quantity kinds, constants, and units.
In ~\cite{bamod_ramanathan2023reasoning}, an ontology called Elementary provides a high-level abstraction of the physical process, that can be linked to components in a technical system.

\subsection{Web of Things and its Relevance to Automation Systems}
\label{sec:wot}

Automation systems are increasingly adopting architectures and technologies behind the Internet of Things (IoT), which is also one of the key drivers behind the idea of creating \textit{digital twins}.
However the general lack of both technical and semantic interoperability in the practical implementations of IoT~\cite{web_wilde2007putting} led the standardization body W3C to put forward a recommendation under the title \textit{Web of Things}\footnote{\url{https://www.w3.org/WoT/}}.
The recommendation includes a schema to \emph{semantically} describe \textit{things} and the interfaces they provide -- this is called the \textit{Thing Description} (TD)\footnote{\url{https://www.w3.org/TR/wot-thing-description/}}.
Listing \ref{lst:td-boiler-01} shows an illustrative example of a digital twin of a boiler modeled as a TD -- later in section \ref{subsec:synth}, we show how knowledge about the physical process this boiler is involved in can be included \textit{in-band} in this TD, and how this enables reasoning about the physical consequences of using the boiler's controls.
In the context of automation systems, a TD, therefore, fits well as a part of the digital twin of a thing~\cite{wot_mayer2017open} because it provides the opportunity to integrate machine-understandable knowledge of its design as well as that of the underlying physical processes the thing is involved in.

\begin{lstlisting}[language=td,numbers=none, caption={An example Thing Description of a boiler}, label={lst:td-boiler-01}]
{ "@context": "<references to ontologies>",
  "title": "Boiler for heating system",
  "@id": "urn:boiler-01",
  "base": "http://blr-01",
  "@type": "brick:Boiler",

  "properties": { "water-out-temp": {
      "title": "Oulet water temperature",
      "@type": "brick:WaterTemperature",
      "readOnly": true, "type": "number",
      "forms": ["href":"/sensors/twout"]}
  },
  "actions": { "fuel-valve-actuation": {
      "title": "Change fuel flow to burner",
      "forms": [{"href":"/actuators/fvalve"}] }
  },
  "events": { "high-temperature-alarm": {
      "title": "High temperature alarm",
      "forms": [{"href":"/alarms/htemp"}]}
  } }
\end{lstlisting}


\section {Approach}

To determine the relevant aspects for the matching of TS to CP, we examined how these are currently described in practice and how this knowledge serves the manual process of matching.
We first summarize our findings and point out that a lack of means to integrate knowledge about physical processes in the machine-understandable versions of design descriptions hinders their automated matching (see Section~\ref{subsec:need}).

The existing fragmentation in machine-understandable knowledge served as a motivation towards building a high-level integrating ontology, which we describe in detail in Section~\ref{subsec:ontology}. Our core contribution here is that we designed an ontology that is based on knowledge from established engineering practices, which allows the workflow for matching to be replicated in an automated manner while reusing descriptions based on available ontologies.

Finally, in Sections~\ref{subsec:synth} and ~\ref{subsec:matching}, we show how the concepts in this ontology can be used to embed the required knowledge directly into the TDs and, further, to create the semantic rules to match them to the CPs.
We make our ontology, it's bridging to existing ontologies, and all examples openly available to the interested readers\footnote{See \url{http://w3id.org/phydit/doc}}.

\subsection{The Need For Knowledge About Physical Processes}
\label{subsec:need}

The design of a TS described in the form of components and their topological relationships (e.g., \textit{part-of} or \textit{connects-to}) is broadly termed as the \textit{system design description} (SDD)~\cite{engineering_dibowski_integrated,sd_nielsen2015systems}.
Knowledge about the physical processes and the mechanisms that the system needs to manage is only selectively described in SDDs because the designer assumes that a human expert reading the SDD will be able to infer them.
Semantic Web ontologies for SDDs also reflect this by providing only taxonomies and relationships to describe components and their topology and do not exploit the opportunity to integrate semantic model of physical processes~\cite{sd_pauen2021tubes,sd_yang2019ontology}.

Many commonly encountered system types (e.g., boilers) have similar technical features and are based on the same or similar physical principles.
For such cases, designers devise reusable CPs that are based on a \textit{generalization} of the system's design: the \textit{abstract system design} (ASD).
This is the CP designer's conception of the TS.
The the relation between the interfaces of a CP, i.e., its IOs, and the sensors and actuators in the ASD are described in text or diagrams.
Finally, in the documentation of the CP, the goals of the program and the expected roles of the sensors and actuator in the physical process are also described.
We use the term \textit{control program description} (CPD) to collectively refer to such explanation (which is currently only in human-readable form). 

Achieving a machine-understandable version of the CPD (to enable automated matching) faces three major gaps.
Firstly, though ASD is a vital part of a CPD, current approaches to semantically describe CPs do not include it as a \textit{first-class abstraction}.
Instead, they only have a provision to annotate the IOs with selective information regarding the physical variables.
We argue that ASDs, because they are in effect \textit{stereotypical} systems, can be semantically described using the same ontologies as for SDDs.
We will show later in this section that a first-class abstraction is indeed required for formulation of generic semantic rules for matching.

Secondly, like SDDs, ASDs need to be linked to the underlying physical process model because this knowledge is required when describing a CP.

Finally, given the possibility of semantically modeling the ASD, concepts to integrate it into the description of the CP are currently missing.

We show that these gaps can be addressed by developing bridging concepts that can interlink knowledge modeled using current ontologies for CPs, SDDs, and physical processes.

In order to determine the particular bridging concepts that are essential for matching, we analyzed descriptions of several existing CPDs---namely, those in the ASHRAE Guideline 36 library\footnote{\url{https://ashrae.org}, machine-readable models are here \url{https://simulationresearch.lbl.gov}} for HVAC functions, boiler combustion control programs from four different manufacturers (described in texts and SAMA diagrams), and a CP library of a large BA manufacturer (described in texts and P\&I diagrams).
In total, 87 different CPDs in the BA domain for applications ranging from air handling units, variable air volume units, ventilation units, chilled water plants, to boiler plants were analyzed.
We also examined exemplary SDDs that were manually matched to the CPD, noting the conceptual aspects that contributed to a successful match. 

Our finding from this study was that amongst the CPDs, some could be matched to SDDs \emph{solely} based on taxonomic and topological features ---for example, in the case of a radiator-based heating CP, it is sufficient to know that the TS should consist of a \emph{radiator} that has a control valve and a temperature sensor.
However, in a large majority (78 of the 87) of the cases, the \emph{description about the physical process} that a CP was conceived for
and the \emph{contextualization of the inputs, outputs, or parameters} in the process 
played an \emph{essential} role in matching it to an SDD.
This demonstrates that the taxonomic and topological knowledge captured by current ontologies for describing TS in SDD and ASD of a CP is not sufficient for the purpose of automated matching.

We could classify the aspects of a CPD that are relevant when reasoning about its match to a TS as 1. \textit{general specification} which describes compatibility of the CP to type of technical system (and in several cases the underlying physical process), 2. \textit{IO specification} that describes the sensors and actuators that their IOs are expected to be connected to, and 3. specification of \textit{design-related parameters} that the CP requires. Table \ref{tab:cpd-to-tsd-mapping} lists these aspects of the CPDs along with examples.

Our study confirms our previous observation that machine-understandable semantic description of CPDs require: 1. an explicit model of the ASD to be a part of the CPD as a first-class abstraction, and, 2. the description of the technical system, both in the form of SDD and ASD, to be linked to description of the physical process (PPD).

We propose to achieve the above through a high-level bridging ontology, which we call PhyDiT (standing for \textbf{Phy}sics-infused \textbf{Di}gital \textbf{T}wins) --- we will refer to concepts in this ontology using the namespace abbreviation of \texttt{pdt}.

\begin{figure}[ht]
\centering
\includegraphics[width=9cm]{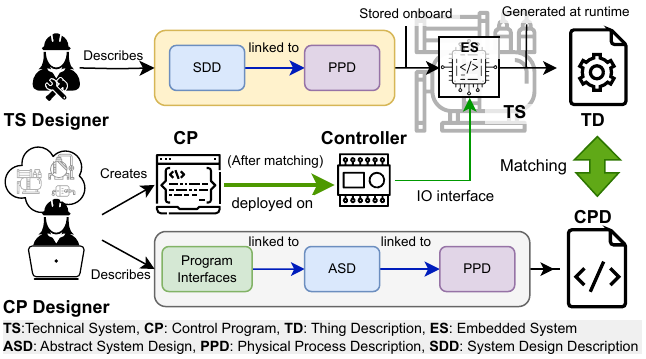}
\caption{Overview of how the integrated knowledge in the two design workflows are used for automated matching.}
\label{fig:workflow}
\end{figure}

Figure~\ref{fig:workflow} illustrates how our proposed approach integrates knowledge in the two design abstractions.

\begin{table*}[t]
\centering
\caption{Summary of how CPs are described in terms of its compatibility to technical systems and physical processes}
\label{tab:cpd-to-tsd-mapping}
\begin{tabular}{|l|l|l|l|}
\hline
\textbf{CP Aspect}     & \textbf{Technical system-related aspect}          & \textbf{Physical process-related aspect}      & \textbf{Example specification}                                          \\ \hline
General                 & Suitable technical system type(s)                         & Physical process and mechanisms               & Boiler with heat generation through combustion                       \\ \hline
Inputs                  & Sensor type and system context                           & Observed variable (quantity kind, etc.)         & Water pressure (in kPa) sensor at pump inlet                  \\ \hline
Outputs                 & Actuator type and system context                         & Manipulated and affected variables            & Fan motor drive (to influence air flow in combustion)                           \\ \hline
Parameters              & Configuration and sizing                          & Limits for variable values              & Boiler capacity, minimum water flow rate          \\ \hline
\end{tabular}
\end{table*}

\subsection{The PyhDiT Ontology}
\label{subsec:ontology}

\begin{figure*}[ht]
\centering
\includegraphics[width=15cm]{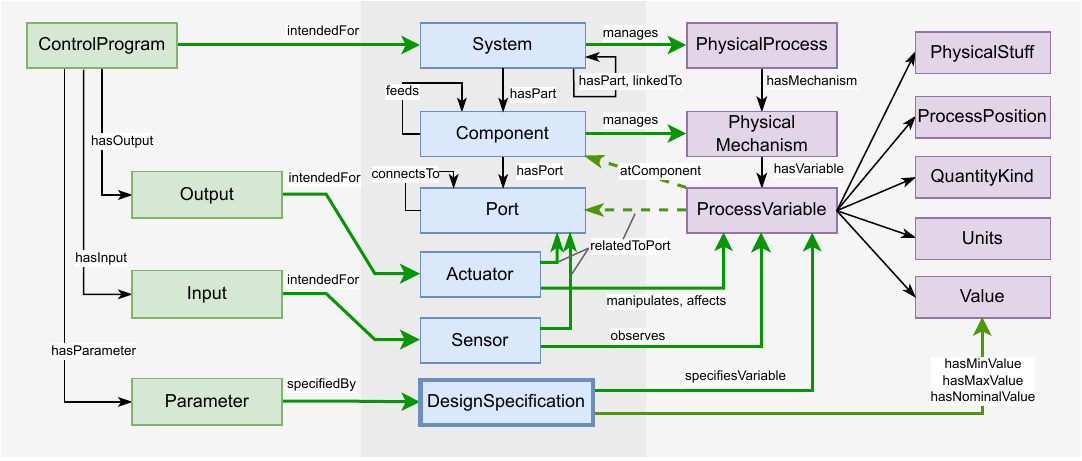}
\caption{Overview of the PhyDiT ontology. The bold lines and borders indicate the bridging concepts we have introduced. The dotted lines indicate relationships that are inferred (e.g., \texttt{atComponent} is inferred if the variable is associated to mechanism that is managed by the component.)}
\label{fig:ontology-overview}
\end{figure*}


\begin{table*}[]
\centering
\caption{Mapping of classes in PhyDit to existing SDD ontologies\\ (*=port is inferred **=no distinct class to model components)}
\label{tab:mapping-to-sddonts}
\begin{tabular}{|l|l|l|l|l|l|l|}
\hline
\textbf{PyDiT Class}  & \textbf{BRICK} & \textbf{ASHRAE 223p} & \textbf{Haystack} & \textbf{RealestateCore} & \textbf{SAREF4BLDG} & \textbf{SOSA-SSN} \\ \hline
System              & System         & System               & system            & Equipment               & Device              & System            \\ \hline
Component           & Equipment      & Equipment            & equip             & Component               & Device**            & System**          \\ \hline
Port                & feeds*         & ConnectionPoint      & -                 & -                       & FlowTerminal        & -                 \\ \hline
Sensor              & Sensor         & Sensor               & input             & Sensor                  & Sensor              & Sensor            \\ \hline
Actuator            & Actuator       & Actuator             & output            & Actuator                & Actuator            & Actuator          \\ \hline
DesignSpecification & -              & -                    & -                 & Document                & Via data properties          & SystemProperty    \\ \hline
\end{tabular}
\end{table*}

\subsubsection{The Required High-level Abstractions}
As high-level abstractions, PhyDiT models three aspects (see Figure \ref{fig:ontology-overview}: 1. The control program and its external interfaces, 2. the technical system in the form of components and its external control interfaces (sensors and actuators), and 3. the physical process in the form of mechanisms and the process variables which represent their state.

We examined whether the classes and relationships within each aspect corresponded well to those in existing concrete ontologies --- this was done to ensure that PhyDiT can be easily bridged to them.
The classes and relationships involved in describing control programs have direct equivalences in CTRLOnt and CDL. Similarly, the model for physical processes corresponds directly to concepts in OntoCAPE, Elementary, Tubes, and PhySys.

There is a much larger variety of ontologies for describing technical systems.
We studied the ontologies that apply to the building systems domain and found that the classes and relationships modeled in PhyDiT can largely be bridged to existing system design ontologies (see Table \ref{tab:mapping-to-sddonts}). 

We now describe the concepts provided by PhdDiT to semantically inter-link the three individual aspects.

\subsubsection{Linking System Design Description to Physical Process Model}
We introduce the relationship \texttt{manages} to link the structural aspects of the technical system and the physical process models.
A \texttt{System} is linked to a \texttt{PhysicalProcess} through the relationship \texttt{manages}. Similarly, a \texttt{Component}, which is a part of a \texttt{System} \texttt{manages} a \texttt{PhysicalMechanism} that is a part of the \texttt{PhysicalProcess}. For example, this can be used to state that a boiler \textit{manages} heat generation process, while its component, the burner, \textit{manages} the mechanism of fuel combustion.

Further, an \texttt{Actuator} can be linked to a \texttt{ProcessVariable} through two possible relationships: 1. \texttt{manipulates}, which signifies the immediate independent variable that the actuation will influence, and 2. \texttt{affects}, which goes through the chain of \texttt{PhysicalMechanism}(s) to list the dependent variables that will be affected.
Similarly, a \texttt{Sensor} can be linked to a \texttt{ProcessVariable} through the relationship \texttt{observes}.
By coupling these relationships to the structural relationships, the \textit{system context} of a sensor or actuator can be inferred in terms of \texttt{Component} and its \texttt{Port} where it can be observed, manipulated, or affected.

\subsubsection{Linking Descriptions of Control Program and System Design}
In the general description of a CPD the kind of technical system and, and optionally, also the kind of underlying physical process for which it is suitable is stated.
In our approach, this specification can be achieved by linking the instance of \texttt{ControlProgram} to the instance of \texttt{System} (that is part of the ASD) using the property \texttt{intendedFor}.
Consequently, the \texttt{PhysicalProcess} and its \texttt{PhysicalMechanism}s can be inferred.

Similarly, by linking sensors and actuators to process variables in the previous step, we have in effect coupled the description of the external interface of the system to the underlying PPD.
Therefore, now by (solely) relating (through \texttt{intendedFor}) the IOs to sensors and actuators in the ASD, the specification of both TS- and physical process-related aspects of the IOs listed in Table \ref{tab:cpd-to-tsd-mapping} are achieved.

Finally, a CP may also require operational parameters that must to be obtained from the design specification of the TS.
For example, the combustion control program may require to know the minimum air flow rate to maintain while operating the blower.
To enable the TS designer to specify such parameters and its values, PhyDiT provides the class \texttt{DesignSpecification}.
In the SDD, an instance of this class can be linked to a \texttt{ProcessVariable} through relation \texttt{specifiedVariable}, and the value can be stated using
the data properties \texttt{hasMinimumValue}, \texttt{hasMaximumValue} and \texttt{hasNominalValue}.
From the CPD's perspective, the need for a design specification to parameterize the CP is expressed by a similarly formulated  \texttt{DesignSpecification} (however without value specification), which is linked to a \texttt{Parameter}.


\begin{table*}[t]
\centering
\caption{Mapping of concepts to generate the TD from SDD.}
\label{tab:td-tsd-mapping}
\begin{tabular}{|l|l|l|l|}
\hline
\textbf{TD abstraction} & \textbf{Metadata}    & \textbf{Mapping to SDD and its semantics}                      & \textbf{Example}             \\ \hline
Thing                   & rdf:type             & Domain class corresponding to the pdt:System instance          & brick:Boiler                 \\ \hline
Thing                   & pdt:manages          & pdt:PhysicalProcess that the TS manages                        & hvac:EnergyConversion        \\ \hline
PropertyAffordance      &                      & Corresponds to an instance of pdt:Sensor in the SDD            & :temperature-sensor           \\ \hline
PropertyAffordance      & pdt:sensor           & Domain class corresponding to pdt:Sensor instance              & brick:TemperatureSensor      \\ \hline
PropertyAffordance      & pdt:component        & The pdt:Component which related to the observed variable       & brick:Burner                 \\ \hline
PropertyAffordance      & pdt:observes         & The pdt:ProcessVariable that the sensor observes               & :water-inlet-temperature      \\ \hline
ActionAffordance        &                      & Corresponds to an instance of pdt:Actuator in the SDD          & :fuel-valve                   \\ \hline
ActionAffordance        & pdt:actuator         & Domain class corresponding to pdt:Actuator                     & brick:ValveActuator           \\ \hline
ActionAffordance        & pdt:manipulates      & The pdt:ProcessVariable that is directly affected              & :fuel-flowrate                \\ \hline
ActionAffordance        & pdt:affects          & The pdt:ProcessVariable that is consequently affected          & :water-outlet-temperature       \\ \hline
ActionAffordance        & pdt:mechanism        & The mechanism that the manipulated variable is associated to  & hvac:Combustion              \\ \hline
ActionAffordance        & pdt:component        & The pdt:Component that manages the above mechanism             & brick:Burner                 \\ \hline
\end{tabular}
\end{table*}

\subsection{Synthesizing TDs}
\label{subsec:synth}

In section \ref{sec:wot}, we highlighted that the computational and network communication capability provided by the ES of a TS is an opportunity to integrate the SDD directly into the TS in form of WoT TD.
We developed a scheme (shown in Table \ref{tab:td-tsd-mapping}) that identifies the required knowledge in the SDD and the abstractions of a TD to which it can be mapped.
In other words, we are infusing the TD with knowledge that is relevant for matching \textit{it} (i.e., the TS) to one or more suitable CPD.
In effect, the TD becomes the \textit{physics-infused digital twin} of the TS.

Listing ~\ref{lst:annotated-td} shows an example snippet of a TD\footnote{The complete example TD is available in the supplementary material} that has been enhanced by knowledge from SDD.
The TD also provides links to the source models.
The \texttt{systemModel} provides the URL of the SDD itself (which could be, for example, stored on-board the ES), while \texttt{simulationModel} can be linked to a resource server from where the FMU of the TS can be downloaded.

\begin{lstlisting}[language=td, caption=The TD of a boiler augmented with knowledge of physical process, numbers=none, label={lst:annotated-td}]
{...
"@type": "brick:Boiler",
"manages": {"@type":"hvac:EnergyConversion"},
"systemModel": "/model/tsd-blr-01.rdf",
"simulationModel":"http://server/boiler.fmu",
"specification": { "min-water-flowrate":
 {"specifiedVariable":{"unit":"qudt:L-PER-MIN",
  "quantityKind":"qudt:VolumeFlowRate", 
  "stuff":"brick:Water"},"hasMinValue": 50}},
"actions": {
  "fuel-valve-actuation": {
    "actuator": {"@type": "brick:Valve"},
    "relatedTo": {"@type":"hvac:Combustion"},
    "manipulates": 
     { "atComponent": {"@type": "hvac:Burner"},
       "stuff":"brick:Fuel",   
       "position": "elem:inlet", 
       "quantityKind":"qudt:VolumeFlowRate"},
    "affects":
     { "atComponent": {"@type":"hvac:BoilerTube"},  
       "stuff":"brick:Water", ...}
},
"properties":{
    "water-outlet-temp":{ 
    "sensor":{"@type": "brick:TemperatureSensor"},
    "observes":
     { "atComponent": {"@type":"hvac:BoilerTube"},  
       "stuff":"brick:Water",..."}}
}
\end{lstlisting}

\subsection{Matching TDs to CPDs}
\label{subsec:matching}
The mappings that we described in the preceding sub-sections allow us to now create formal rules for matching a TD ($\mathcal{T})$) to a CPD, i.e. an instance of \texttt{ControlProgram} ($\mathcal{C})$).
We define $\tau_x = C(x)$ as a function that returns the OWL class of an RDF individual $x$. The predicate $subClassOf(\tau_{sub},\tau_{super})$ is true if $\tau_{sub} \sqsubseteq \tau_{super}$.

For mapping $\mathcal{T})$ with $\mathcal{C})$, we define three formal rules to validate (i) Type Compatibility, (ii) IO Compatibility, and (iii) Parameter Availability.
We wish to emphasize that the rules proposed below are meant to illustrate the conceptual benefit of our approach --- the exact formulation may be adapted to suit the application- or domain-specific needs.
\newline

\textbf{Type Compatibility}  A CP can be specified as being suitable for one or more types of TS (e.g., a temperature control program can be stated to be suitable for both heating and cooling systems). We also found from our study that a CP that has been developed for a more specific TS type can also be used with the general type.
Therefore, the following rule checks if a TS type conceived by the CP designer (in the ASD) is the same or sub-type of the TS type $\tau_t = C(\mathcal{T})$ in the SDD. 
\small
\begin{flalign*}
& \exists s_c (System(s_c) \land intendedFor(\mathcal{C},s_c) \land &\\
& subClassOf(C(s_c),\tau_t)) \rightarrow isSystemCompatible(\mathcal{C},\mathcal{T})&
\end{flalign*}
\normalsize
Similarly, to match that the type of physical process(es) conceived for the CP is the same or sub-type of that the TS instance manages:
\small
\begin{flalign*}
& \exists p_c,s_c ((System(s_c)\land intendedFor(\mathcal{C},s_c) \land manages(s_c,p_c)) &\\
&\rightarrow \exists p_t (manages(\mathcal{T},p_t) \land subClassOf(C(p_c),C(p_t)) &\\  
&  \rightarrow isProcessCompatible(\mathcal{T},\mathcal{C})&
\end{flalign*}
\normalsize

\textbf{IO Compatibility}
In principle, here, descriptions of the CPs' IOs are matched to the TD's interaction affordances.
For every \texttt{Input} interface of a CP, a match is sought to a TD \texttt{PropertyAffordance}, and similarly, \texttt{Output} are matched to \texttt{ActionAffordance}.
The inference rules check if the sensor or actuator type related to the TD's interaction affordance is same or sub-class of the sensor or actuator in the CPD, and if the corresponding variables that is stated as being observed, manipulated, or affected are \textit{equivalent}.
We define a function $equivalent(x,y)$, which returns true if the variables $x$ and $y$ have the same stuff, quantity, quantity kind, and process position.
We do not compare the units; we use the quantity kind to evaluate the semantic compatibility.
This function can be extended to also compare the component or the physical mechanism to which the variable is related.
\small
\begin{flalign*}
&\forall i (Input(i) \land hasInput(\mathcal{C},i) \land intendedFor(i,s) &\\
&\land observes(s,v_c) \rightarrow \exists a (hasPropertyAffordance(\mathcal{T},a) &\\
&\land oberves(a,v_t) \land sensor(a,s_t) \land equivalent(v_c,v_t))) &\\
&\land subClassOf(C(s), C(s_t))\rightarrow isInputCompatible(\mathcal{C}, \mathcal{T})&
\end{flalign*}
\begin{flalign*}
&\forall o (Output(o) \land hasOutput(\mathcal{C},o) \land intendedFor(o,r) &\\
&\land manipulates(r,v_{mc}) \land affects(r,v_{ac}) \rightarrow &\\
&\exists a (hasActionAffordance(\mathcal{T},a) \land manipulates(a,v_{mt}) &\\
&\land affects(a,v_{at}) \land subClassOf(C(r), C(a)) &\\
&\land equivalent(v_{mc},v_{mt}) \land equivalent(v_{ac},v_{at}))) &\\
&\rightarrow isOutputCompatible(\mathcal{C}, \mathcal{T})&
\end{flalign*}
\normalsize
Further, similar to the above rule, we define $hasActionCoverage(\mathcal{T}, \mathcal{C})$ to be true if each action affordance of the $\mathcal{T}$ has a compatible output in $\mathcal{C}$.
This rule will ensure that we do not match a CP only based on a subset of the TS actuation interfaces.
\newline

\textbf{Parameter Availability}
To check if a \texttt{Design\-Speci\-fication} required by a CP is available in the TD, we compare the equivalence of the variables each specifies:
\small
\begin{flalign*}
& \forall p (hasParmeter(\mathcal{C},p) \land specifiedBy(p,d_c)&\\ 
&\land specifiesVariable(d_c,v_c)\rightarrow \exists d_t(specification(\mathcal{T},d_t)&\\
&\land specifiesVariable(d_t,v_t) \land equivalent(v_c,v_t))&\\
&\rightarrow isSpecCompatible(\mathcal{T},\mathcal{C})&
\end{flalign*}
\normalsize

Given a specific $\mathcal{T} $and a specific $\mathcal{C}$, compatibility is inferred through the conjunction of the above rules.

\section{Evaluation}
\label{sec:evaluation}
Our evaluation aimed to verify if SDDs and CPDs, each originating from independent designers (and created using different ontologies), could be matched based on the high-level concepts provided by PhyDiT.
Additionally, instead of having a centralized knowledge graph or even using the complete SDD of a TS, we wished to infuse the required knowledge in the form of metadata in the TD so that it is self-contained.
Therefore, we want to confirm if the approach of embedding the selective parts of the SDD into the TD is sufficient for matching to CPDs in real-life applications.

To evaluate our approach, we chose a real-life setup of office buildings where thirty-four air-handling units (AHUs) were installed and commissioned.
No two AHUs were exactly the same --- they differed in their topology and the type of components involved.
We created the OWL version of the SDDs for the thirty-four AHUs using the Haystack ontology, while using Tubes to describe the underlying physical process.

The AHUs were already commissioned using one of eight choices of CPs meant for AHUs available in the BA manufacturer's CP library.
This means that even if two AHUs differed in terms of component types and topology, the same CP could fulfill the automation requirement.
From the commissioning documents, we knew the engineers' choice of CP for each AHU. From the project logs, we estimated that, on average, it takes an hour (varying between 20 and 90 minutes) to read and understand an SDD and search for a suitable CP in the library

The eight CPs were each described using text and diagrams. In principle, they contained the ASD of the AHU that the CP was suitable for along with the specifications of the sensors and actuators required to be connected to the programs' IO.
Based on such descriptions, we created corresponding CPDs in OWL using the BRICK and Elementary ontology.
Additionally, we also created CPDs for applications that are functionally close to AHUs --- namely, for ventilation, fan-coil, and variable air volume units.
This is because even though two CPs may match from physical mechanism and IO perspective, the sequence of operation and timing of control actions often differ based on the type of the technical system.

We chose the above-described setup because it presented the essential characteristics to generalize our evaluations --- i.e., a heterogeneous system described using different ontologies.

The thirty-four AHU instances were each equipped with an ES in which a Java-based program was deployed that could synthesize a TD from SDD according to the mappings identified in Table \ref{tab:td-tsd-mapping}.
This program can work with any SDD based on an ontology that is bridged to PhyDiT. 
Through the ES's network interface, an AHU's TD could be obtained at run-time.

We developed another Java-based program, which, given the network address of a TS, retrieves its TD and then uses SPARQL queries corresponding to the rules listed in section ~\ref{subsec:matching} to find the CPD(s) that matches the TD.
\newline

\textbf{Results}
We compared the matches suggested by our program with the choice of CPs that were deployed (and tested) at the installation.

For our 34 AHUs, 31 ($>$90\%) were correctly matched; one was matched to two CPs, and two were not matched to any CP.

For one of the AHUs, two CPs were (correctly) suggested as a match, though only one was suitable from the process dynamics perspective (in this case, a cascade control loop was required for temperature control). For the two remaining AHUs, our approach did not match any CP because the physical context of an actuator did not conform to any description in the CP library. In this case, the SDDs had a valve in the outlet of the heat exchanger instead of in the inlet, which is what the CPDs in our library specified.

To address multiple matches, we intend to develop a method to specify and match requirements regarding process dynamics, and further, a way to employ simulation models (which can be linked to the TDs) for verifying the chosen CP.

To address the case of the AHU, where no match was found, we intend to develop a method to specify design options/variants in the ASD.

\begin{table}[]
\caption{The role of TD metadata in enabling a successful match. The grouping (A-D, with the number of SDDs shown in brackets) shows how they occur together.}
\label{tab:role-of-metadata}
\begin{tabular}{|l|cccc|}
\hline
\multirow{2}{*}{\textbf{\begin{tabular}[c]{@{}l@{}}TD metadata and the percentage of \\ matches where they were necessary\end{tabular}}} & \multicolumn{4}{c|}{\textbf{\begin{tabular}[c]{@{}c@{}}Grouping according to  \\the metadata necessary\\ (marked with $\bullet$)\end{tabular}}}                                                                        \\ \cline{2-5} 
& \multicolumn{1}{c|}{\textbf{\begin{tabular}[c]{@{}c@{}}A\\ (4)\end{tabular}}} & \multicolumn{1}{c|}{\textbf{\begin{tabular}[c]{@{}c@{}}B\\ (8)\end{tabular}}} & \multicolumn{1}{c|}{\textbf{\begin{tabular}[c]{@{}c@{}}C\\ (16)\end{tabular}}} & \textbf{\begin{tabular}[c]{@{}c@{}}D\\ (4)\end{tabular}} \\ \hline
General: System type (100\%)                                                                                                             & \multicolumn{1}{c|}{$\bullet$}                                                        & \multicolumn{1}{c|}{$\bullet$}                                                        & \multicolumn{1}{c|}{$\bullet$}                                                         & $\bullet$                                                        \\ \hline
General: Physical process type (88\%)                                                                                                    & \multicolumn{1}{c|}{}                                                         & \multicolumn{1}{c|}{$\bullet$}                                                        & \multicolumn{1}{c|}{$\bullet$}                                                         & $\bullet$                                                        \\ \hline
Property: Sensor type (37\%)                                                                                                             & \multicolumn{1}{c|}{$\bullet$}                                                        & \multicolumn{1}{c|}{$\bullet$}                                                        & \multicolumn{1}{c|}{}                                                          &                                                          \\ \hline
Property: Observed variable (62\%)                                                                                                       & \multicolumn{1}{c|}{}                                                         & \multicolumn{1}{c|}{}                                                         & \multicolumn{1}{c|}{$\bullet$}                                                         & $\bullet$                                                        \\ \hline
Property: Component and position (12\%)                                                                                                  & \multicolumn{1}{c|}{}                                                         & \multicolumn{1}{c|}{}                                                         & \multicolumn{1}{c|}{}                                                          & $\bullet$                                                        \\ \hline
Action: Actuator type (100\%)                                                                                                            & \multicolumn{1}{c|}{$\bullet$}                                                        & \multicolumn{1}{c|}{$\bullet$}                                                        & \multicolumn{1}{c|}{$\bullet$}                                                         & $\bullet$                                                        \\ \hline
Action: Manipulated variable (62\%)                                                                                                      & \multicolumn{1}{c|}{}                                                         & \multicolumn{1}{c|}{}                                                         & \multicolumn{1}{c|}{$\bullet$}                                                         & $\bullet$                                                        \\ \hline
Action: Affected variable (62\%)                                                                                                         & \multicolumn{1}{c|}{}                                                         & \multicolumn{1}{c|}{}                                                         & \multicolumn{1}{c|}{$\bullet$}                                                         & $\bullet$                                                        \\ \hline
Action: Component and position (62\%)                                                                                                    & \multicolumn{1}{c|}{}                                                         & \multicolumn{1}{c|}{}                                                         & \multicolumn{1}{c|}{$\bullet$}                                                         & $\bullet$                                                        \\ \hline
General: Design parameters (12\%)                                                                                                        & \multicolumn{1}{c|}{}                                                         & \multicolumn{1}{c|}{}                                                         & \multicolumn{1}{c|}{}                                                          & $\bullet$                                                        \\ \hline
\end{tabular}
\end{table}

We furthermore evaluated the relevance of the different aspects (e.g., system type, sensor type, etc.) of the TD metadata for matching performance. This was done by selectively removing aspects and verifying whether the implementation still produced a correct match.
Across all 32 successful matches, Table~\ref{tab:role-of-metadata} shows the relevance of each aspect that we embedded in the TDs of the AHUs towards enabling successful matching.
%
%
We grouped the SDDs according to the set of aspects that were required to correctly match them. Table \ref{tab:role-of-metadata} shows these groups, the number of AHUs in each, and the metadata that was necessary: In group A, only knowledge about the system type, sensor type, and actuator type is required for correctly matching---illustrating that for these matches, no knowledge about the ``elaborate'' physics is required. On the other hand, without knowledge about the observed, manipulated, and affected physical process variables, 20 AHUs ($>$62\%) would have been matched incorrectly.
In groups C and D, we see that describing the observed physical variable makes the sensor type information redundant (but not vice versa), which shows that semantics related to the physical process play a more fundamental role in describing the CP's IOs.

Through our evaluation, we also confirm that the knowledge required for matching could be embedded into the TDs, thereby achieving a decentralized (and dynamically adaptable) method of making system knowledge accessible.
\newline

\textbf{Limitations}
Apart from the need to model the specification of process dynamics and design variants of TS that we recognized in our evaluation, we also recognize that our approach currently does not include the dimensions of system requirements and goals.
Though the requirements and goals are often implicit in the inter-linked descriptions of the technical system and the physical processes that it manages, an explicit integration of semantic description of requirements (for e.g., through goal-oriented requirements engineering approach~\cite{re_gore_van2001goal}) will enable formal verification that the matched CP fulfills the process goals. We are currently researching this possibility.

\section{Conclusion}
Our proposed approach of using the bridging concepts in the PhyDiT ontology to link ASD and SDD to PPD showed demonstrable benefit in enabling automated matching of a TS to a CP.
The ontology also enabled the extraction and embedding of knowledge of physical mechanisms into the TDs, which serve as a distributed knowledge base.

The PhyDiT ontology could be easily bridged to existing engineering ontologies, so semantic rules and software program implementations based on it are not tightly coupled to a specific choice of ontologies.
The high-level concepts also make the rules for TD metadata extraction and matching flexible for modifications and extensions to suit specific domains.

Our approach shows the concrete benefits of infusing knowledge of the underlying physical mechanisms into digital twins, which will open up further avenues to explore and achieve autonomous automation.

\bibliographystyle{splncs04} 
\bibliography{references} 

\end{document}